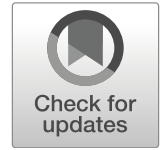

# COVID-19 what have we learned? The rise of social machines and connected devices in pandemic management following the concepts of predictive, preventive and personalized medicine

Petar Radanliev [1] · David De Roure [1] · Rob Walton [1] · Max Van Kleek [2] · Rafael Mantilla Montalvo [3] · Omar Santos [3] · La'Treall Maddox [3] · Stacy Cannady [3]

## Abstract

**Objectives** Review, compare and critically assess digital technology responses to the COVID-19 pandemic around the world. The specific point of interest in this research is on predictive, preventive and personalized interoperable digital healthcare solutions. This point is supported by failures from the past, where the separate design of digital health solutions has led to lack of interoperability. Hence, this review paper investigates the integration of predictive, preventive and personalized interoperable digital healthcare systems. The second point of interest is the use of new mass surveillance technologies to feed personal data from health professionals to governments, without any comprehensive studies that determine if such new technologies and data policies would address the pandemic crisis.

**Method** This is a review paper. Two approaches were used: A comprehensive bibliographic review with R statistical methods of the COVID-19 pandemic in PubMed literature and Web of Science Core Collection, supported with Google Scholar search. In addition, a case study review of emerging new approaches in different regions, using medical literature, academic literature, news articles and other reliable data sources.

**Results** Most countries' digital responses involve big data analytics, integration of national health insurance databases, tracing travel history from individual's location databases, code scanning and individual's online reporting. Public responses of mistrust about privacy data misuse differ across countries, depending on the chosen public communication strategy. We propose predictive, preventive and personalized solutions for pandemic management, based on social machines and connected devices.

**Solutions** The proposed predictive, preventive and personalized solutions are based on the integration of IoT data, wearable device data, mobile apps data and individual data inputs from registered users, operating as a social machine with strong security and privacy protocols. We present solutions that would enable much greater speed in future responses. These solutions are enabled by the social aspect of human-computer interactions (social machines) and the increased connectivity of humans and devices (Internet of Things).

**Conclusion** Inadequate data for risk assessment on speed and urgency of COVID-19, combined with increased globalization of human society, led to the rapid spread of COVID-19. Despite an abundance of digital methods that could be used in slowing or stopping COVID-19 and future pandemics, the world remains unprepared, and lessons have not been learned from previous cases of pandemics. We present a summary of predictive, preventive and personalized digital methods that could be deployed fast to help with the COVID-19 and future pandemics.

✉ Petar Radanliev
petar.radanliev@eng.ox.ac.uk; petar.radanliev@oerc.ox.ac.uk

1 Department of Engineering Sciences, University of Oxford, Oxford, UK

2 Department of Computer Science, University of Oxford, Oxford, UK

3 Cisco Research Centre, Research Triangle Park, Durham, NC, USA

 

## Introduction

With the rise of COVID-19, most western governments seem determined to design apps and surveillance mechanisms as a means of response. During the first wave of the COVID-19 pandemic, these proved successful in some countries including Taiwan, South Korea and China. Western countries are developing similar monitoring and surveillance approaches, hoping to replicate that success and prevent significant damage from rising risks of a second wave. With monitoring as the main drive, we could be limiting the potential for pandemic management. Designing separate systems in isolation could lead to the realization that such systems are coupled by default. But by being designed in isolation, such systems become proprietary and not interoperable with similar systems built by others. This can also result in the repeat of past cybersecurity mistakes. We investigate interoperability in predictive, preventive and personalized medical system design, which could become the topic of concern when upgrading the same systems we are building today, for pandemic management at the latter stages of this outbreak or future outbreaks.

### The case of COVID-19

COVID-19 was declared as an outbreak in January 2020 by the World Health Organization (WHO). Within 6 weeks, it became a pandemic, and by mid-March 2020, the COVID-19 pandemic had generated 24 times more cases than the severe acute respiratory syndrome (SARS) outbreak [1]. One of the differences is that the amount of digital data produced in the time period prior to and during SARS in 2003 is produced within minutes today in 2020 [1]. Big data has enabled geospatial mapping of COVID-19 for epidemic transition analysis, but challenges remain in 'data aggregation, knowledge discovery, and dynamic expression' [2]. There is also an opinion that many countries did not learn from the previous two coronavirus pandemics (Middle East respiratory syndrome (MERS) and SARS), resulting in an inadequate risk assessment of the urgency and the speed of the pandemic [3]. Taiwan on the other hand has learned from the SARS pandemic and has developed a fast response health mechanism that recognized the crisis and activated emergency measures to contain the outbreak [4].

### Digital solutions for the COVID-19 outbreak

This review paper is looking for digital solutions that enhance the speed of response during pandemics. Fast information flow is needed for understanding the dynamics and development of an epidemic and to aid prevention, control, and decision-making and for informing actions. This research studies interoperability of digital health solutions and the benefits from integration of these systems. A lack of data in the initial stages of the COVID-19 outbreak led to its initial rapid spread across the world. Despite the abundance of digital solutions for preventing future pandemics, the world remains unprepared. A specific focus in this study is the speed of response enabled by the social aspect of human-computer interactions (e.g. social machines) and the increased connectivity of humans and devices (e.g. IoT).

### The role of connected devices and social networks

The Internet of Things (IoT) refers to technologies with sensing, networking, computing, information processing and intelligent control capabilities. IoT is commonly used to describe various smart objects and things, with computational abilities, that exchange information, potentially even on a global level, perceiving the world as one adaptive and interconnected system [5]. This communication and exchange of information is done via different protocols including WAN, LAN/LoRaWAN, Bluetooth, Z-Wave, INSTEON, ZigBee, Wi-Fi, 4G and 5G. When social networks are integrated with the IoT, a new concept emerges, named Social Internet of Things (SIoT). The SIoT potential architecture, policy and network structure have already been tested with simulations [6]. These simulations show that SIoT depends on the type of relationship (e.g.) Ownership object relationship (OOR); Social object relationship (SOR); Co-work object relationship (C-WOR); Co-location object relationship (C-LOR); and Dynamic Social Structure of Things (DSSoT). The 'Social Collaborative Internet of Things' (SCIoT) goes even further in respect that social objects are grouped and collaborate by interacting and sharing [7], in other words, working as cognitive coupled systems. These concepts emerge as values in the advancements of smart healthcare monitoring and management systems that are using big data analytics, such as iHealth and m-Health [8]. The diversity of SIoT implementations points to problems of scale and interoperability in terms of aggregation of data for responding to a pandemic.

## Methodology

This is a review paper, investigating different approaches taken across the world in managing the COVID-19 pandemic with digital technologies. To forecast events, digital technologies currently use artificial intelligence (AI) to process large amounts of data. However, AI applications require narrowly defined use cases and large training datasets to 'teach' the app how to forecast. This would result in a slow adaptability of AI to managing a new and fast-changing pandemic. To use AI in pandemic management, social machines and connected devices will require a different approach to AI training, for example, selective re-purposing of AI systems already trained for other uses, like temperature measurement. A social machine is defined as 'an environment comprising humans and technology interacting and producing outputs or action which would not be possible without both parties present'.[1]

As we have seen with most mobile app concepts for pandemic management, they depend on human and technology interaction *at scale*, that is, with mass participation by large segments of the population. To achieve the targets of the UK's digital contact programme, for instance, it was anticipated that a minimum adoption rate of 60%[2] of the UK population was needed.[3] However, where such apps require the capture and sharing of sensitive data with various entities such as the government, potential users can become wary and reluctant to participate. One study found that over 70% citizens were unwilling to download contact tracing apps.[4] Since this data is already shared by many humans on social media and mobile apps, this opens up questions on how to design a privacy preserving contact tracing app that ensures that a sufficient percentage of humans will download the pandemic management app. To investigate this, we refer to social science methods and relate the concept of social machines, with digital health and pandemic management. An additional approach is to use standard open-source intelligence (OSINT) to gather data that is available in the public domain.[5] This includes searches to public repositories of Internet-exposed and vulnerable IoT devices (for instance, shodan.io).

## Social machines in digital health for pandemic management

Global pandemic management requires fast data and knowledge acquisition with real-time visualization of results. This can be achieved by interactive spatial transmission and social management of data in geographic information systems [9]. This relates to the 'Social Internet of Things' (SIoT), which integrates people and smart devices interacting within a social structure of IoT, often through IoT platforms [10, 11]. The 'Social Collaborative Internet of Things' (SCIoT) represents a more advanced form of SIoT where social objects collaborate, interact and share information, creating cyber-physical ecosystems [12].

Social machines are human-machine communities on the web (e.g. Wikipedia). These often change, like living hybrid organisms, and may be represented as cyber-physical and socio-technical systems [13]. Such systems can be observed through a plurality of archetypal narratives [14]. The SIoT utilizes users' needs, preferences, social drives and the surrounding environment to generate situation-aware services [15]. This refers to cognitive reasoning within temporary social structures, combining users, objects and services in the form of a 'Dynamic Social Structure of Things' [15]. Cognitive collaboration between connected devices can be achieved by combining machine learning with middleware technologies [12]. Some studies even suggest the creation of emotional models for human-machine interactions [16, 17], as expressed in Table 1. Such human-machine interactions could provide valuable feedback for pandemic monitoring apps.

The social networks described in Table 1 have large numbers of participants and can provide answers to complex problems [6]. Hence, when designing pandemic monitoring apps, we should focus on the value of this human-computer interaction for large numbers of participants and a wide range of devices. An approach to managing new mass surveillance technology that links personal data would be to enable the consolidation of connected devices into virtual objects for remote access and control. This could create a dynamic virtual network connecting different domains and facilitating sharing of resources in a cloud environment [18]. This would allow us to initiate the development of a truly global response and monitoring of pandemics.

## Social collaborative Internet of Things and artificial intelligence

The SIoT opens new opportunities, such as fast and diverse application development though a dynamic end-to-end connection between devices. For example, there is already a consensus among Chinese experts that Internet of Things (IoT) can help in diagnosis and treatment of the pandemic [9]. The SIoT enhances the IoT with improved intelligence, context-awareness and cognitive reasoning [15], such as an IoT cognitive hierarchy thinking mechanism [19]. Some include using social network theory for social relations of mobile nodes in IoT [20]. If we consider the concept of mechatronics, where functionality is transferred from the machine system into an information systems [21], then the SIoT already represents a form of cognitive engine. Cognition in the form of an aware-SIoT that perceives time, space and activity, depends on building social relationships between 'things' [20].

Cognition in SIoT could resolve many social issues. For example, one such issue is incorrect self-diagnosis through personal digital health data. This creates increased and sometimes unsustainable demand for medical practitioners. At best such information could be used by clinicians for understanding a patient's lifestyle [22]. Cognition in health social machines could prevent the damage of self-diagnosis and instead provide tailored interventions by de-coupling activity data into a consolidated longitudinal archive for an integrated approach

**Table 1**    Evolution of the Social Collaborative Internet of Things

| The evolution of Social Collaborative Internet of Things | | | | |
|---|---|---|---|---|
| Humans and things provide reactive and proactive re-time data | Social Collaborative Internet of Things | H2H, T2T, and Human to thing | Human: consumer and producer Thing: consumer and producer | Analysing the need of humans using the information generated by humans and things in collaboration with other humans and things |
| Things provide reactive data; humans provide reactive data | Social Internet of Things | H2H, T2T | Human: consumer Thing: consumer and producer | Sensing the local and global environment using information from things and humans and collaborating with other things and humans |
| Things provide reactive and proactive data | Internet of Things | T2T | Human: consumer Thing: producer | Sensing the local environment using local information |

*Pervasiveness – availability* (vertical axis, left)
Sociality – connectivity (horizontal axis, bottom)

to 'intervention, management, mitigation and sense-making' [22], supported with FAIR stewardship for distributed data analytics [23]. With the advancements in machine moral mind [24], and criteria to evaluate the health of a social machine [25], the interactions with social machines could evolve into a meaningful communication based on moral agency, where humans act as a 'cognitive anchor', noting that even humans need models for mental behaviour [24].

## Social collaborative Internet of Things and quality of experience in digital healthcare

The SIoT operates in a way that is similar to how humans interact with social networks. The physical social connections between humans and things represent the baseline for the logical configurations of social communities, involving humans and things [17]. The logical configuration adopts individual social network behaviours into a universal social network of all entities [17]. The IoT contains many aspects of social networking, where value emerges from the exchanges between users and devices, enabling service that would be of more interest because of interactive filtering (e.g. embedded systems gathering and analysing data before communicating to social networks of people and devices), to provide better quality of experience [17].

Whilst it is likely that during rapidly unfolding crises, aspects other than service design will be given priority in the design and deployment of new digital interventions, it is well established that ease of use and user experience, combined with a perceived value for participants, could increase uptake and adherence, resulting effectiveness of such interventions. Moreover, since the effectiveness of such interventions depends on high rates of user adoption, rapid uptake and participation are crucial. Thus, systems designed for gathering personal data during pandemics should provide value and quality of service to encourage more participants. Participants could also be encouraged by different systems of trust. There are various schemes of trust, with diverse trust scaling, dimensions, inferences and semantic meanings [26]. For example, for trust in social networks, Amazon and eBay use star ratings; P2P networks measure quality of downloaded files and speed.

Some medical devices are considered IoT devices. There are many challenges identifying what closed or open source software is running in medical devices and other IoT devices in the industry. The list of software used in any device (including non-IoT devices) is referred to a software bill of materials (SBOM). Vendors and consumers lack awareness of not only the software components they are using and adopting but also the underlying security vulnerabilities that affect such software. The National Telecommunications and Information Administration (part of the United States Department of Commerce) is leading an effort to provide guidelines to producers and consumers of technology on how to create effective SBOMs that can help identify the presence of known security vulnerabilities.[6] This effort is divided into four major working groups: framing; awareness and adoption; formats and tooling; and healthcare proof of concept.

The healthcare proof-of-concept working group created a report that provides information on SBOM creation by the MDMs; consumption by the healthcare delivery organizations (HDO); and the exercising of the procurement, asset management, risk management and vulnerability management use cases. Challenges and areas for improvement were identified and documented around an absence of naming standards, partial version information, vulnerability vs exploitation qualification, configuration vulnerabilities, no authoritative end of life database and lack of patching level data.[7]

## Pandemics and socio-economic hardships

COVID-19 has triggered a significant shock in the economy with social distancing and quarantine policies triggering economic recession. Economic hardship could harden the impact of pandemics, and digital technologies provide some solutions for increasing economic value during global pandemics. While current focus is on pandemic management, with the increased focus on digital management of the COVID-19, significant focus is also placed on connected devices and IoT. Prior to the pandemic, it was estimated that IoT will create a value in 2020 projected at 1.9 trillion USD [27] with connected IoT devices reaching 20.8 billion in 2020 [28]. These connected devices and social networks enable data collection and exchange that was not possible during previous pandemics, and this represents a different healthcare business ecosystem. Healthcare business models need to adapt accordingly to secure ethical access to this new value.

Some of the areas of focus to secure ethical access to this new value are in monitoring and control, big data for business analytics, information sharing and collaboration [29], among many other value areas. Harnessing the IoT potential also requires a digital strategy; for example, the cloud offers value in high-scale real-time analytics of big data, and IoT middleware offers lightweight real-time analytics [30]. Cloud IoT also enables a vision for worldwide IoT integration [31]. Apart from the medical and business side, there is also a social value from IoT social connection and networking, which can be explored by combining social computing with opportunistic IoT, for analysing the physical and digital social communities simultaneously [32]. By adding intelligence in objects, smart objects collect data, interact and control the physical world while symbiotically sharing data and being interconnected with other things [33], though a diverse set of communication technologies that enable remote management [33].

## Cyber risk from digital solutions in pandemic management

With the integration of large numbers of social IoTs in a digital healthcare design, a large data distribution platform can be subjected to a relatively simple large-scale attack, triggering a huge economic cost and even loss of life [34]. These emerging risks from IoT in healthcare ecosystems need to be understood in the digital solutions design process. One way to understand these risks is through existing reference architectures for attack surface analysis [35]. Cloud IoT ecosystems are seen as pervasive and disruptive, enabling many different applications that are gaining momentum, but with many issues surrounding privacy and security [36]. There are however existing cyber risk assessment approaches for smart systems [37], for IoT systems [38] and for cloud provider environments [39]. The use of these could enhance the privacy of digital heath solutions. SIoT is in the same risk category. SIoT objects can anonymously establish communications and cooperate opportunistically with neighbouring IoT devices, but trust management in SIoT is a concern [7]. Therefore, existing performance metrics for security operations analysis [40] and failsafe recommendations [41] should be consulted in the design of digital heath solutions.

## Case study of digital solutions in pandemic management

Two data sources were used extensively in case studies: (1) the Google Scholar search engine and (2) a critical appraisal of Publons' new COVID-19 publication index that investigates sound scientific practices of preprints and papers. As COVID-19 is a global and fast-spreading pandemic, as visualized in Fig. 1, the case study research is not focused on investigating individual region responses in great detail. We designed the diagram in Fig. 1, with data from this study, using the 'draw.io' open access web software.[8] The different lines represent the spread of COVID-19 regionally, nationally and internationally, leading to a global pandemic.

The COVID-19 pandemic has spread globally (demonstrated conceptually in Fig. 1, without reliance on concrete data) even before most governments realized the need to act; for example, at the time of writing, the first known case in France was in December 2019, almost a month before the French government announced the first case in France.[9]

---

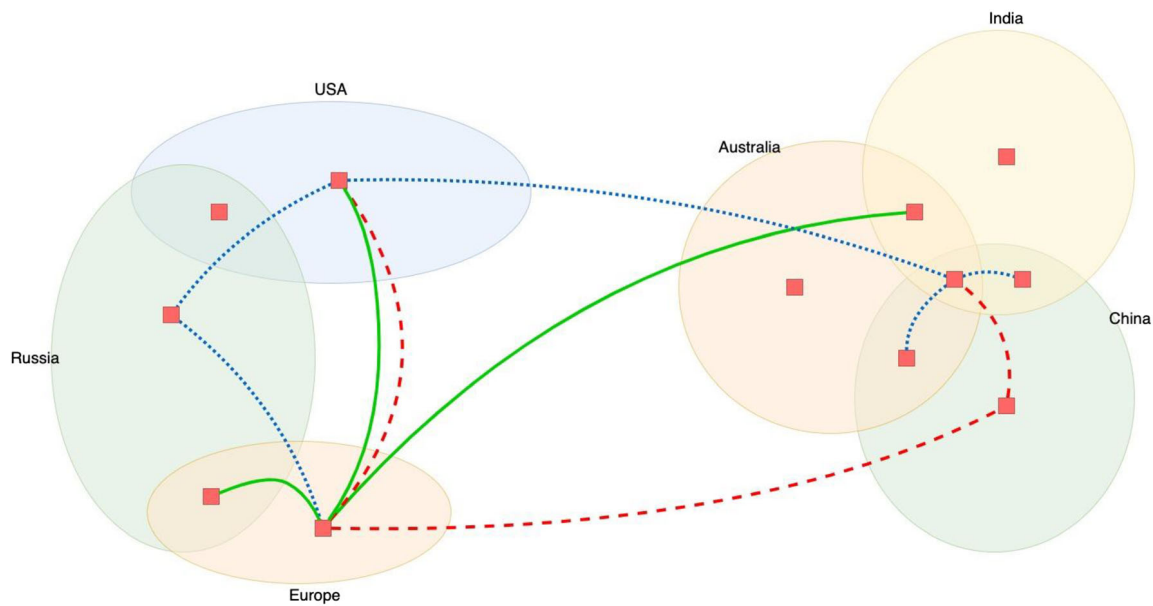

**Fig. 1** Pandemic as an epidemic on a national or global level

Therefore, studying regional approaches in detail at this stage, when we do not truly understand the data, would probably not produce the results that we require. Instead, focus was placed on the regions that performed best during the initial outbreak, with the objective to gather and accumulate knowledge on the effectiveness of different methods and with the aim of making recommendations for replicating this success in the second wave.

## The case of COVID-19 in Taiwan – Role of social machines in digital democracy

The Taiwanese culture of civic technology, a fusion of technology, activism and civic participation, created a strong response to COVID-19 [42]. The vTaiwan digital democracy platform [43] presents a decentralized community of participants. The platform hosts social machines that uncover a problem and then help participants to understand this problem, even while the situation is changing daily or hourly. Taiwan implemented actions (interventions) with great speed through the platform [43]. Some of the actions were aimed at reducing panic buying, such as the application programming interface (API) that provided real-time, location-specific data to the public on critical items' availability. To design the interactive face mask maps, for example, the digital minister worked with hacktivists through a digital chatroom.

A second example of success was a bottom-up information sharing platform on which individuals created a strong participatory collective action for sharing real-time reports about symptoms using social smartphone apps, call-in lines and other approaches. The results were integrated with location history by community-created apps, enabling individuals to track and compare their exposure. This approach supported proactive behaviour, where people who were concerned about being exposed had a tool to determine if they should self-isolate prior to developing symptoms. The authorities also had a tool to monitor people in self-isolation with a mobile phone–based 'electronic fence' that traces mobile phone signals to track location and follow up with two phone calls every day to make sure people stay home. The Taiwanese system sends alerts to police if self-quarantined people leave their home, and the police would contact or visit the person within 15 min of the alarm being triggered. The Taiwanese 'electronic fence' system could be the first in the world designed for this purpose.[10]

The success factor for such large-scale social coordination was the common-sense approach and the privacy protection of participating individuals. This platform reduced economic impact greatly, through successful containment that avoided social distancing measures. People either self-isolated, avoided or disinfected contaminated locations, depending on their location history review. The community data enabled more targeted responses that have proven more effective than in all other economies, including the USA, the EU and China, at least in the first wave. The case of Taiwan has proven that social inputs and coordination can prove more effective in resolving complex fast-evolving problems than centrally run top-down approaches that take a while to deliver and, in some regions, lack buy in by the population. Further, the community-driven development of software tools in this case was proven to be fast and precise, driven by society participation.

---

[10] https://www.reuters.com/article/us-health-coronavirus-taiwan-surveillanc/taiwans-new-electronic-fence-for-quarantines-leads-wave-of-virus-monitoring-idUSKBN2170SK

This review paper is focused on the human-computer interactions in pandemic management, where technology creates value when built into social machines, where a large number of participants adding input in mobile apps. However, it is worth mentioning that AI and ML can also add value in purely digital approaches to solving problems. AI/ML applications generally work best with narrowly defined use cases accompanied by large, hand-built training datasets to 'teach' an application how to execute against its use case. This results in slow adaptability of AI/ML to new opportunities, unless an existing capability can be re-purposed. For example, consider the Kogniz Health product[11]—a stand-alone infrared digital camera with an integrated ML trained to identify passers-by who show signs of fever and send alerts in real time via a variety of standard messaging tools. The ML capability is possibly re-purposed from an existing ML use case to monitor and alert on temperature variations beyond acceptable limits in an industrial setting. This approach could work in advanced economies because of partially existing infrastructure, strong know how and technological/financial strength. But it is difficult to see how these tools can be made operational in developing countries, at the required speed.

## The case of COVID-19 in the UK

Given the lack of treatment such as a vaccine for COVID-19, the world's current focus is on global surveillance, and IoT and other new technologies including thermal cameras or IoT sensors [44], and face recognition [45], that can play a crucial part in mapping the spread of infection, and in detecting disease early as part of smart systems with near real-time data tracking and alerts [3]. Currently IoT-enabled drones are used for public surveillance, enforcing quarantines and even for faster disinfecting with agricultural drones. IoT can be used for enforcing compliance of infected people in quarantine, by monitoring and tracing potential infections caused by breach in quarantine. The speed and scale of IoT connectivity could enhance the UK's National Health Service (NHS) ability to monitor high-risk patients in home quarantine. This is currently done with healthcare workers going door-to-door, posing a risk to them, and using up resources for home checks that could turn out to be unnecessary. IoT real-time live cameras could be used for patient checks, cloud databases could be used for patients to upload their temperature measured on home thermometer, and drones could be used to check temperature with infrared cameras. The aggregated IoT mobile data from infected patients can be used for tracing people who might have been in contact with infected patients. The IoT is already used for remote monitoring of in-house patients with serious conditions. Self-reporting apps such as the 'C-19 COVID symptom tracker' enable quick buildup of maps and other data that support people to prevent infection and medical professionals to plan their response. One of the main benefits from app design that operates as a social machine is the speed of deployment and data collection. For example, the UK 'Covid Symptom Tracker' app reached 700,000 downloads and app registrations in 24 h [46]. It is almost impossible to gather such large datasets with any of the traditional data collection tools. But societal participation in data collection, during times of crises, seems to be producing relevant data at a very fast rate.

Earlier attempts were different in approach. For example, a mobile phone application called FluPhone App [47] used Bluetooth and Wi-Fi signals to track human interactions and asked users to report symptoms. The data collected was used for understanding how diseases spread in complex human network structures, using information such as how frequently people meet and how much time they spend together. One of the main weaknesses in this approach is that smartphones can accurately determine location between 7 and 13 m [48], and COVID-19 spreads in people that are within less that 2-m proximity [49]. Tracing Bluetooth and Wi-Fi location also does not always help with mapping objects in movement (e.g. flights, trains and buses).

## The case of COVID-19 in China – Role of IoT in social machines

The Chinese approach for compulsory registration of phone numbers to an app, prior to using public transport, and allocating colour codes through social media apps, could be more effective in using big data for automated forecasting of contagion risk, but it triggers privacy concerns [1]. The Chinese approach is enabled through the wallet app Alipay and the social media, messaging and mobile payment app WeChat. The apps display a code that can be read by smartphones: green means unrestricted and yellow or red implies a need to self-quarantine or be in a supervised quarantine. Codes are based on tracing location history to determine contacts with outpatients or people hospitalized in the past 14 days and are refreshed every day at midnight. Many restaurants, coffee shops and shopping malls require to see a green code to enter. But there are already concerns about glitches and incorrect codes that ban people from travelling or entering public spaces,

While the Chinese approach for digital epidemic control has triggered privacy concerns and mistrust, similar public response can be expected in the EU and USA where people have even lower level of trust in their governments [1]. This mistrust can be fuelled even more with the increased attacks against individuals working from home during the pandemic.[12] There is also a big difference in how such cyber-attacks

---

are mitigated. One malware sample we discovered was a wiper designed to destroy infected systems. The malware was aimed at people working from home but also at healthcare organizations. The filename '冠状病毒.exe' which translates to 'coronavirus' uses several techniques to delete data from both the file system and registry in an attempt to disrupt system operations. The Taiwanese approach has on the other hand caused far less mistrust despite involving big data analytics, the integration of national health insurance databases, access to travel history from individual's location databases, code scanning and online reporting. The main difference between the EU, the USA, China and Taiwan on the other hand is the difference in public communication strategies [1].

### The case of COVID-19 in South Korea – Social machines and IoT in smart cities

South Korea was one of the countries where COVID-19 started early in the global outbreak, and their experience with SARS has proven effective in lowering the death rate through strict quarantine and fast testing. South Korea deployed a tracking app called Corona100m (Co100), informing people of known cases within 100 m of their location.[13] Further tracking measures included a 'smart city technology system' that uses surveillance camera footage and credit cards transactions to track past movement of infected patients.[14] These tools are in addition to the Coronamap[15] website tracing the location history of the infected and the information search engine Coronaita[16] designed specifically for finding information on infected areas.

### The role of tech companies in COVID-19 pandemic management

The Chinese response to COVID-19 has shown that information systems for pandemic management, in terms of database and spatial analysis and mapping, can be designed and made operational fast but is limited by the commercial software that is in existence [2]. Some of the main challenges are the large-scale people flow globally, the long incubation period of this virus, lack of symptoms in some of the infected, virus reoccurrence after recovery, evolving nature of the virus targeting different DNAs and the different level of impact on individuals. If pandemic tracking information system is designed to operate regionally, its functions are limited in global pandemic prevention monitoring.

### Statistical bibliometric analysis of scientific research records

Two different software tools were heavily used for a bibliographic review of the types of data used in pandemic management: (1) R Studio was used with the 'bibliometrix' [50] package to analyse the large sets of data records, and (2) a computer programme for bibliometric mapping called VOSviewer [51] was used for graphically representing large data records in a bibliometric map. The Web of Science Core Collection was identified as the data source for the two different software tools. The data search parameters were narrowed to search for articles on COVID-19 in 2020. Then, a historic search parameter was used to identify methods applied in research on previous pandemics.

From the first search result, we searched for 'COVID-19', and out of 2264 records, we used the Web of Science filter and downloaded the most relevant 500 records. When analysed with the R 'bibliometrix' package, we selected two most striking figures. In Fig. 2, we can see a three-fields plot of keywords, topics and countries.

To compare the three-fields plot analysis with alternative mapping of scientific research on COVID-19, we designed a graph by corresponding author countries Fig. 3.

To compare these results, with the entire 2254 records on the Web of Science Core Collection, we used the Web of Science result analysis tool and found similar results as seen in Fig. 4.

Although in Figs. 2 and 3, China is shown as the dominating country in the scientific research on COVID-19, we compared the three-fields plot analysis, by using the same data file of records and designed a collaboration network as a social structure Fig. 5. What is clear from Fig. 5 is that the UK is performing strongest in collaborative research on COVID-19, and this collaboration is strongly represented between the UK, the USA and Canada.

We continued this analysis by doing a historical search on the Web of Science Core Collection for scientific research on global pandemics. This resulted with 3708 records, and we used the Web of Science tool for research analysis (see Fig. 6). The historic analysis showed that countries that have led the scientific research on global pandemics in the past are not represented as strongly as China in the scientific research on COVID-19. Since China was first affected in this global pandemic, we could speculate that regional research effort is driven by the effect of the pandemic. It is possible that other countries would pick up and start producing more scientific research, but such comparison can only be confirmed when data becomes available. At present, these graphs represent statistical analysis of the current data, as a snapshot in time, compared with historical analysis in Fig. 6.

What is concerning from the results of this statistical analysis of scientific research is the extremely low record numbers in scientific research on pandemics. When we compare the 3708 historical records on global pandemics on the Web of Science Core Collection, with 372,454 results on business, 573,901 results on art and 455,356 records on fishing, then the problem becomes concerning. The world simply does not seem prepared for COVID-19, not to mention a deadlier global pandemic. Similar conclusions are presented in our review of new forms of digital data used in pandemic management.

---

[13] https://www.theguardian.com/commentisfree/2020/mar/20/south-korea-rapid-intrusive-measures-Covid-19#maincontent
[14] https://www.smartcitiesworld.net/news/news/south-korea-to-step-up-online-coronavirus-tracking-5109
[15] https://coronamap.site/
[16] https://coronaita.com/#/

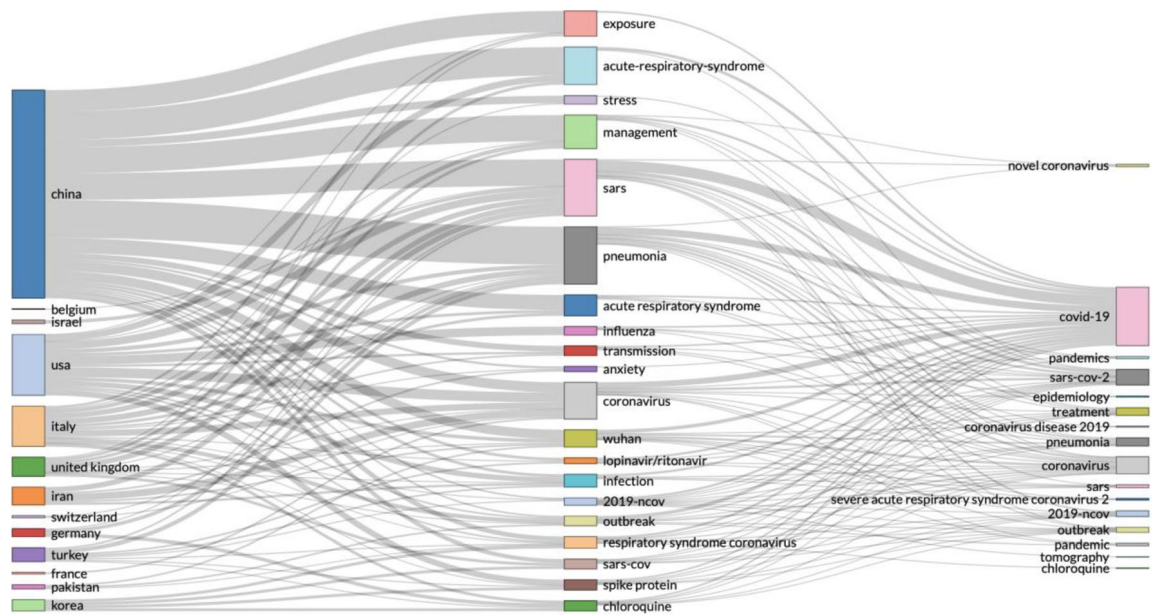

**Fig. 2** Results of a statistical bibliometric analysis of the 500 most relevant research records on COVID-19

## Types of digital data in pandemic management

In this section we detail our search for scientific research using digital data. Firstly, such data does exist. Digital healthcare (e.g. e-Health, m-Health, Telemedicine) includes management of large, heterogeneous, distributed, multiscale and multimodal datasets distributed across edge-fog-cloud healthcare paradigms [52]. Complementing such paradigms, smart cities generate vast amounts of data, captured and stored in different heterogenous formats that need to be transformed to be of analytical value [53]. IoT and big data analytics are emerging trends in new biomedical and healthcare technologies [54]. These various different types of data need to be integrated to ensure fast and effective response during pandemic management. Although there are concerns about data privacy, there are many types of data that are already shared openly in different community settings (e.g. skill set, goods sold, vehicles movement). Some personal data needs to be evaluated with specific privacy requirements in consideration, but there is also a lot of data shared in open community marketplaces [55].

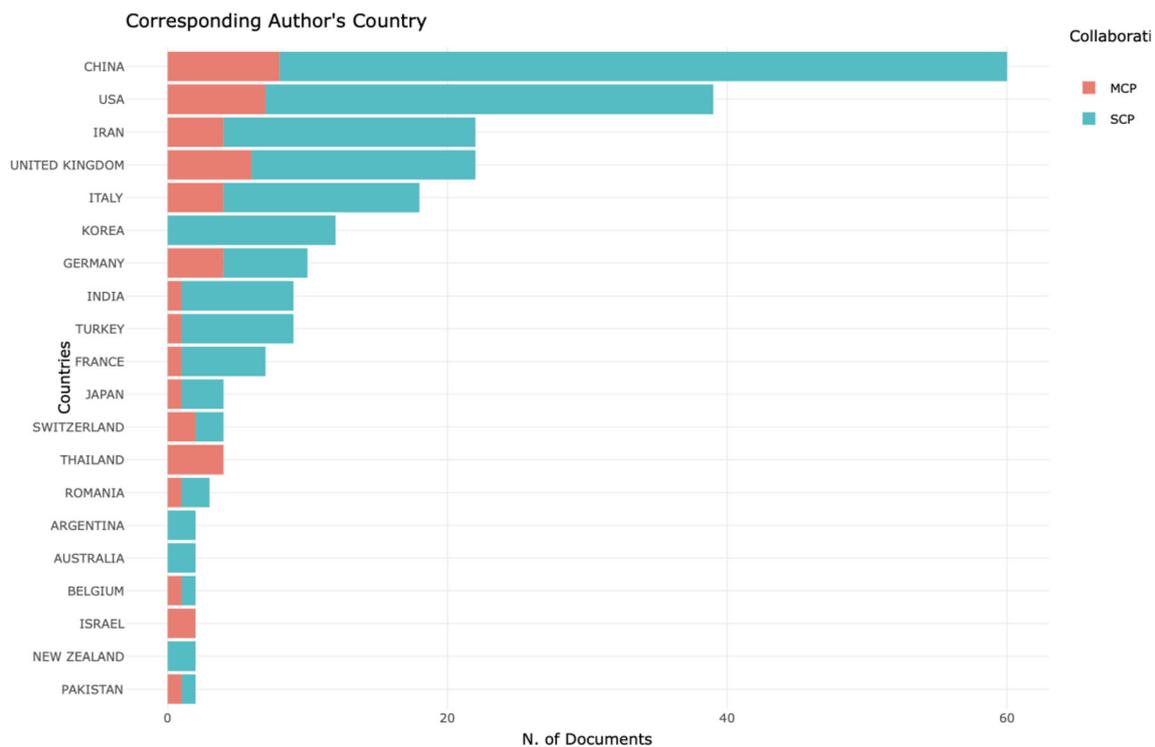

**Fig. 3** Corresponding author by country—COVID-19 research in 2020

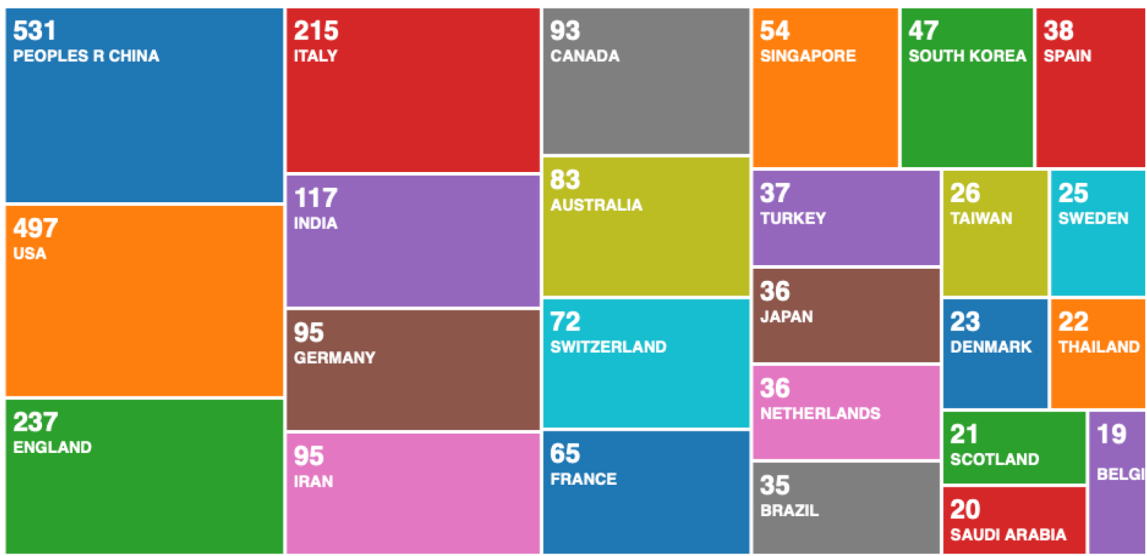

**Fig. 4** Web of Science—results analysis tool

**Fig. 5** Collaboration network as a social structure of COVID-19 scientific research records in 2020

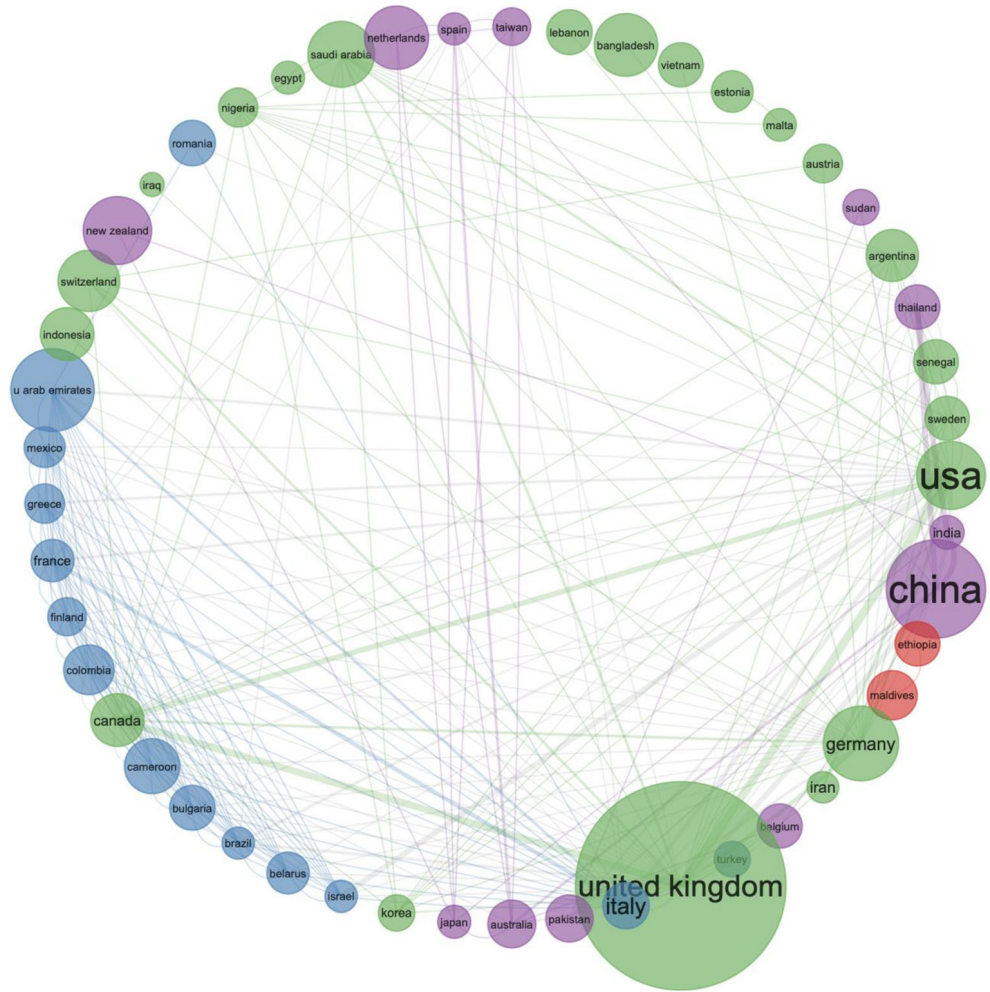

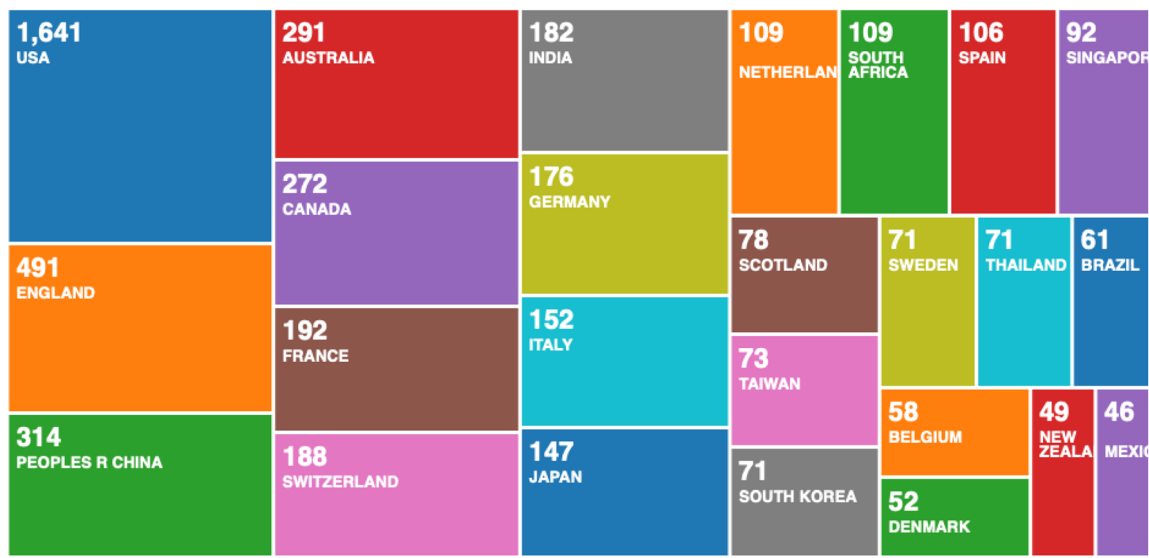

**Fig. 6** Historic analysis of scientific research on global pandemics 1900–2020

Focusing on one specific data type (e.g. big data) would not necessarily result in low latency. Adding to this, the peer-reviewed scientific research on individual data sources is still in its infancy, and the COVID-19 pandemic is already causing deaths globally. Currently, there are only five scientific publications on the Web of Science Core Collection under the search on 'COVID-19 and big data'. We reviewed all five publications, and only one actually discussed big data solutions. This one publication is only a viewpoint, not a research article. A similar search on 'pandemics and big data', using historical records since 1900 (to 2020), produced only 59 records. On the other hand, search results on 'big data' alone produced 88,410 records. It is unclear if this result is caused by lack of funding or lack of interest in the past on this research topic. Hence, we continued with different search parameters, and here we discuss few different types of data that need to be integrated in the pandemic management analytical processes, but the scientific research on these topics is very limited.

Time-stamped data refers to data that includes the event time (the time of the event) and processed time (the time data was processed). By recording the time of capture and time of collection, data analytics could recreate location journey, by replaying people's steps and even derive findings on estimating or predicting best future actions. Our search on the Web of Science Core Collection, on the topic of 'COVID-19 and time-stamped data', resulted in no results, while the search of historic records, from 1900 to 2020, on 'pandemic and time-stamped data', resulted in only 3 records.

Genomics data that involves analysing the DNA of patients could identify new drugs and improve healthcare with personalized treatment. But converting genomic data into value is challenging, because of the large volume of data that requires significant processing resources, combined with the need to unify diverse teams from bioinformatics data specialists, to clinical specialists on the front line. Nevertheless, significant progress has already been made. Sequencing the first genome took longer than a decade. With current tools, it takes just 2 days despite these limitations in handling large data. Despite these advancements, our search on the Web of Science Core Collection, on the topic of 'COVID-19 and genomics data', resulted in 0 results, while the search of historic records, from 1900 to 2020, on 'pandemic and genomics data', resulted with 43 records. A similar search on 'COVID-19 and high-dimensional data' also resulted in no records, and the search of historic records, from 1900 to 2020, on 'pandemic and high-dimensional data' resulted in only two records. However, to quote a famous aphorism frequently used in this context [56], the 'absence of evidence is not the same as evidence of absence'. We are aware that the data is currently incomplete so we cannot assume for sure that the research is not happening. We will run this analysis again in a year and see how it has changed and how it compares to previous pandemics.

Healthcare data contains a large number of interrelated variables (e.g. weight, blood pressure, cholesterol levels). To investigate the connection between intraday physical activity and sleep, actigraphy data was applied to a Bayesian framework for function-on-scalars regression [57], resulting in many proposed predictors. Similarly, it is possible to check an individual for the expression of thousands of genes, resulting in millions of possible gene combinations, some of which may be predictors. This is described as high-dimensional data, where the number of features exceeds the number of observations.

To observe high-dimensional data for pandemic management, one would need to analyse multiple measured and recorded parameters including, for example, immune system status, genetic background, diagnosed diseases, operations, treatments, blood analysis, nutrition and alcohol-tobacco drug consumption. Open sharing of such personal medical data would require people to trust in technologies such as the secure multiparty computation (also known as privacy-preserving computation)[17] and differential privacy.[18] Such technologies could enable a broader interpretation, based on trust in privacy preserving technologies. More work needs to be done on identifying and promoting the benefits for private individuals to start sharing openly medical data. Genetic data and personal biodata privacy are evaluated with a unique moral obligation for protecting patients' dignity, and there are ethical limits on how such data can be shared [58].

Analytic transactional (transact and analyse) data comes in handy in resolving this issue in the healthcare environment. Analytic transactional data enables shifting real-time processing with new metrics and in-memory computing. Instead of using one technology to run applications, and separate technology to run batch analysis, creating linked layers and a patched system, analytic transactional data analysis uses a single layer for transactions and analytics. This approach would enable monitoring health indicators in real time while performing other functions simultaneously. Despite its scientific usefulness, the search on 'COVID-19 and analytic transactional data' and the search of historic records, from 1900 to 2020, on 'pandemic and analytic transactional data' both resulted in no records. The search was repeated with multiple terms (e.g. 'analytic transactional', and 'translytic' among other terms).

Spatiotemporal data can be analysed for building an interactive view of data points in similar and different regions, to analyse data from localized areas or to map and compare spatial distribution in individual regions [59]. In previous pandemics, spatiotemporal analysis has been used to identify cholera hotspots in Zambia [60]. With COVID-19 spreading so fast, the scientific peer-review process is struggling to catch up. Currently, there is only one scientific publication on the Web of Science Core Collection under the search on 'COVID-19 and spatiotemporal data'. On the other hand, there are only 24 records on the Web of Science Core Collection under the search on 'pandemic and spatiotemporal data', which means that scientific research is lacking on this topic.

Since our search of established peer-reviewed scientific research on the Web of Science Core Collection resulted with too few records to conduct statistical analysis, in the next section, we instead applied a quantitative case study research methodology. In the case study research, we used other search engines, such as Google Scholar. However, the consistently low records on the Web of Science Core Collection present an overwhelmingly strong argument that highly ranked scientific research, in a sufficient quantity for statistical review and analysis, is currently missing.

## Summary of digital solutions for COVID-19 and future pandemic management

The current digital health infrastructure requires enhancements to support medical efforts in COVID-19 and future pandemics management. Such enhancements are not only required in the field of artificial intelligence algorithms but also in digital infrastructure for data collection in relation to current and past location monitoring, disease transmission forecasting and health monitoring. This is similar to enhancements of infrastructure as response to other natural disasters (e.g. floods, earthquakes). Fortunately, digital technologies have advanced significantly since the last pandemic, presenting a variety of digital solutions that could be adopted for pandemic management relatively easy.

The COVID-19 crisis also changed the perception of medical data sharing, leading to wide acceptance of new mobile apps designed for various pandemic management functions, all based on some form of sharing personal data. Just a few months ago, it would have been unprecedented to even think about one million downloads for a mobile app that requires sharing your personal medical data, including current and past location history. There seems to be a consensus among governments about the benefits of such mobile apps. There are, however, some measures that are operational in individual regions, but have not even been considered in others. In Fig. 7, these tools and measures are integrated, to design a more comprehensive data collection and migration, for epidemic monitoring of regional risk trends and indices of high-risk people or areas. We designed this diagram with data from this study, using the 'draw.io' open access web software.[19]

In the past decade, there was a fast innovation of new digital technologies. This resulted in the new concept of digital health, on which some of the solutions in Fig. 7 are based. Although these digital solutions are already operational and can easily be adopted for pandemic management, without proper planning significant challenges could emerge. Some challenges could relate to capturing duplicate data, hence adding significant workload on medical workers that collect,

---

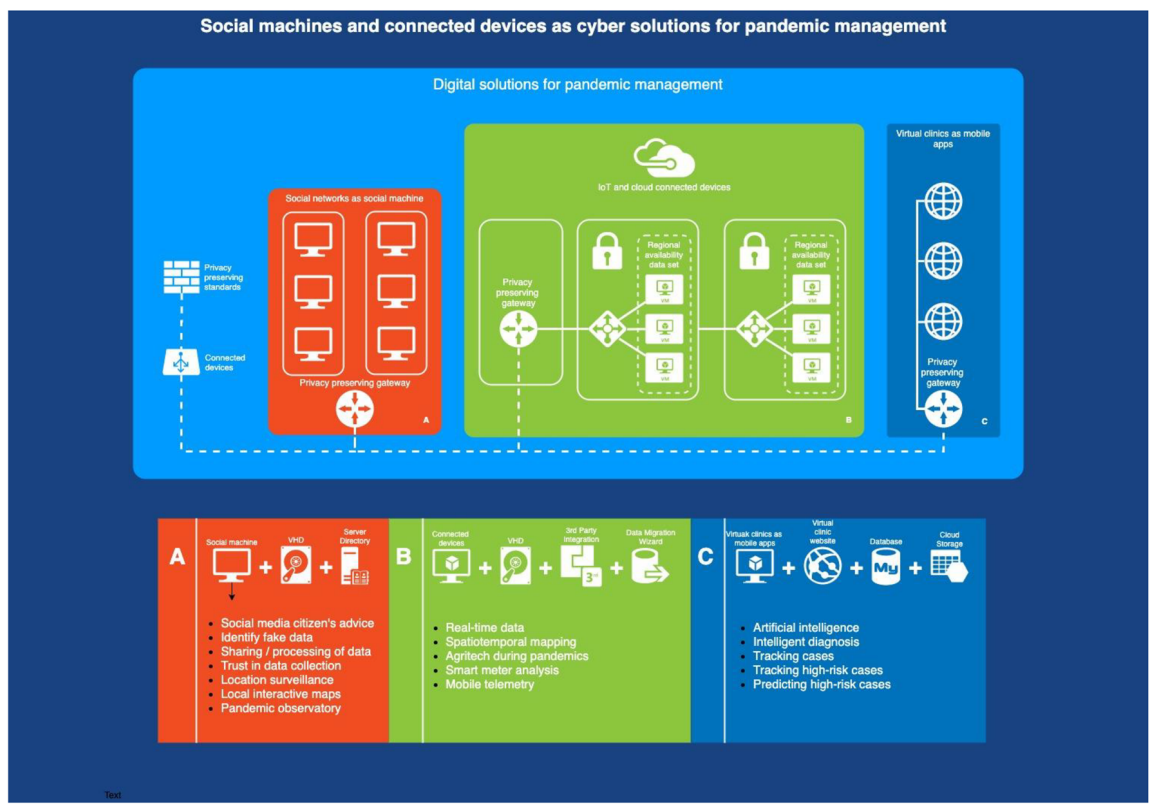

**Fig. 7** Digital solutions for pandemic management

manage and use the data. The design of these digital health solutions separately could lead to lack of interoperability when their integration is crucial for making evidence-based decisions. Individual design could also lead to complexity, for example, hindering the ability of medical professionals to use digital health systems. Open standards for data structure and interoperability in this discipline would help.

While pandemics require a global response, not all countries have the capacity to govern the epidemic response in line with privacy preserving policies. Technologically advanced nations have the advantage in this field and could be faster to use this leverage to advance the global digital health management. Technologically advanced nations could expand these interoperable data systems and tools, strengthening data governance and building capacity to collect and interpret data. However, digitalisation on its own does not improve the health outcome. A coordinated effort is needed that includes both public and private sectors. Past cases have proven that Zika and dengue outbreaks can be predicted within 400 m, months in advance and with 88.7% accuracy, while in Vietnam, an electronic surveillance platform is operative for 44 communicable diseases and syndromes in 63 provinces and 711 districts.[20] These examples confirm that interoperable digital health systems lead to faster detection and response and increased preparedness. We review in more detail these interoperable digital health systems as shown in Fig. 7.

## Case study on social networks operating as social machines in digital healthcare

Social networks with large numbers of participants can provide answers to complex problems [6]. The virtualisation of IoT devices into virtual objects for remote access and control creates a dynamic virtual network connecting different domains, facilitating sharing of resources in a cloud environment [18]. This creates opportunities for fast and diverse application development through a dynamic end-to-end connection between devices that include those following.

### Social media citizen's advice

A social media citizen's advice type system could be designed to connect people with skills and expertise to those who need free access to the advice. This could include legal advice, health, well-being and nutritional advice, technical home computing advice, DIY and small-scale problem-solving engineering advice. This could link to a new app in a format allowing talking about points of concern and should include an option for sharing symptoms and predicted evolution.

---

### Identify fake data

Social media should start classifying and flagging misinformation relating to virus spread; and healthcare systems that monitor these social networks or that directly collect reports from patients should include automated analysis to identify postings concerning symptoms. Fake data in digital systems based on human input is a well-studied subject [61], and digital health systems should be based on existing recommendations.

### Seamless global sharing and processing of geographical data that preserves privacy

The complex and diverse set of data sharing restrictions imposed by countries, regions and companies, sometimes embodied as laws, standards or policies, need to evolve into internationally supported and consistently standardized protocols for the seamless sharing of data, especially where data is kept private for geo-political or commercial reasons [44]. Such data sharing would enable the development of a large database that might be enhanced with artificial intelligence tools for processing data for early detection and better virus containment.

### Definitions of trust in data collection and analysis during pandemics

In some countries, such as the EU and the USA, people have low level of trust in their governments [1]. Hence, gaining trust in digital systems for data collection and analysis during pandemics is important. Trust is described differently in social networks; for example, Amazon and eBay use star ratings, while P2P networks measure quality of downloaded files and speed. There are various schemes of trust, with diverse trust scaling, dimensions, inferences and semantic meanings [26].

### Population location activity surveillance

A mobile phone app can be used to track down the people who have been in close contact with those diagnosed with COVID-19. An app could use contact information (such as calendar entries/social nets) to track potential infections. In addition, cloud, IoT and social media can be used in combination with a new app, for general population location activity surveillance of quarantined people (by locality) and to design worldwide interactive maps. The cloud enables a vision for worldwide IoT integration [31]. Anonymised IoT medical data sharing though the cloud enables the development of interactive maps of COVID-19 cases worldwide. The cloud and IoT technologies can be used for image or video tracking of (super)spreaders or those regularly flouting social distancing advice. The phone app could enable linking to volunteering groups that allows (verified/tested) people who have got through COVID-19 to be part of an immune volunteer resource pool. In the development of pandemic management interactive maps, the definitions of trust, privacy and security in digital systems for data collection and analysis during pandemics need to be based on globally accepted online standards.

### Local interactive maps

A localized COVID-19 mapping system can help residents monitor total cases in their areas, which will enhance value of data sharing and strengthen uptake of government measures.

### COVID-19 pandemic observatory

A pandemic observatory can be used for COVID-19 and for fast response in future pandemics. The observatory would be a cloud-based platform designed to ingest, store, share, analyse and visualize COVID-19-related datasets, e.g. to support and integrate data collected through an app that people voluntarily install to allow access and tracking of (a) health and wellbeing details; (b) access to location data; and (c) to insert infection data for mapping behaviour with timeline and geography of the virus (findings might be incubation time; mapping of first symptoms; length of feeling sick; and response to identification of sickness).

## Case study on connected devices operating as social machines in digital healthcare

### Anonymised real-time data from wearable devices for spatiotemporal mapping and healthcare

Wearable devices and other types of healthy sensors measuring heartbeat, blood pressure and body temperature are abundantly present in many urban areas. Future design of such wearables (e.g. Fitbits, Apple Watches) can be adapted to enable body temperature and other symptoms that would enhance early self-diagnosis of viral infections. Allowing anonymised data to be collected, analysed and shared in real time via the cloud can be used for building spatiotemporal mapping and for enhancing the idea of 'connected healthcare', without breaching data privacy and security [44], by following existing advise on end-user control [62] and automation control systems [63]. Future wearables should be encouraged to include skin sensors for temperature monitoring, which would be linked to geo-mapping and local medical or volunteering teams to ensure people are safe and secure. Wearable devices should include automated hand washing assessment on washing hands properly. Wearable sensors can be used to quantify lung volume and functions, or point of care systems could provide care at home and alleviate the pressure on NHS. Respiratory parameters can be estimated using wearable sensors, spirometers, activity sensors and SPO2 sensors to provide early detection and improved prognosis in patients. The goal should be to create value [64], and it is the design that drives value from connected devices [65].

### Agritech during pandemics

Agritech is the use of technology in agriculture, horticulture and aquaculture for improving yield, efficiency and profitability.[21] During global pandemics, agricultural harvest faces a significant drop in workers.[22] Agritech can reduce inputs and increase yields, but adoption is costly, complex and slow. COVID-19 is an opportunity for governments to support big and small agritech companies and new pioneering start-ups[23] to help rapidly accelerate it, not only to help the agriculture industry to survive COVID-19 but also to prepare the industry and develop the technology in anticipations of future global pandemics.

### Smart meter analysis

Geographical analysis of electricity, water and gas consumption may be used to evaluate adherence to social distancing; with mobile alerts or warnings sent to non-adhering areas. Privacy-preserving measures could include focus on adherence over a large geographical area and not on individual households. Similar analysis is already published on the domestic energy usage patterns during social distancing.[24]

### Mobile telemetry for lockdown monitoring

Mobile phone networks hold personal mobile phone telemetry that can be used as a real-time data source for feeding AI techniques to identify and send message alerts to those likely to be ignoring the lockdown. Personal mobile phone telemetry can also be used in an app to accompany contact tracing though Bluetooth and Wi-Fi data in identifying those who have come into contact with someone who has COVID-19. In addition, some smart home systems can help with pandemic management. Existing supervised intrusion detection systems [66] could be adopted to detect home visitors that should be in self-isolation.

### mHealth and virtual clinics as mobile applications

Mobile health (mHealth)[25] is not a new concept. It involves using mobile devices, usually through a secure mobile app, for healthcare management, data sharing and improving patient experience. Apart from mHealth, there are also virtual clinics that can provide personalized medical advice and encourage participatory surveillance for geo-located predictions. Currently operational virtual clinics include PingAn in China.[26] They are effective at reducing costs and saving time and may provide access to medicines through machine dispensers. In developing countries, the cost and time required to build, equip and distribute new machine dispensers may be prohibitive. Adapting the already built and distributed existing food and drink dispensing machines could be faster and more effective. A simple restocking with non-prescription over-the-counter medicines while changing the prices would provide a fast and cost-effective solution. Virtual clinic pharmacy testing kit outcomes should be linked to a central database and any pandemic monitoring system. Virtual clinics should also enable video and supervised tutorials for the self-management (at home) of oxygen therapy, including protocol, sensor use (SPO2) and O2 generation/administration.

To be effective in pandemics when most of the population is in self-isolation, virtual clinics should evolve into mobile applications enabling home access. Similarly to the South Korean 'self-quarantine safety protection' app that traces GPS location to track if people are avoiding quarantine and enables self-quarantined people to communicate with medical caseworkers,[27] teleconferencing, or video conferencing, has long been used in telehealth,[28] also known as telemedicine. It involves a real-time video call, while exchanging other medical data (e.g. blood pressure) required for a medical check. Collaboration apps for virtual meetings are also well established in the medical profession. Collaboration apps enable remote data collection from patients and analysis by medical practitioners based in different geographical locations.

Global pandemics have exposed a weakness in the personal protective equipment (PPE) manufacturing and supply chains. Some hospitals have resorted to the use of substandard PPE because they had nothing else to use. Individual GPs have reported using the black market to obtain PPE at a much higher cost.[29] Urgent PPE deliveries have also been found of substandard quality and had to be taken out of use.[30] Advice has also been given to medical staff to refuse treating patients if they do not have adequate PPE.[31,32] Given these widespread concerns about the ability to manufacture and supply sufficient PPE in the face of sudden demand, we should be looking at how robotic technologies could be used to improve the safety of medical professionals during a second wave or in future pandemics. There are three types of healthcare robotics: medical, assistive and rehabilitative. When AI enhanced, robots could attend to multiple patients in quarantined areas, for example, by delivering drugs. Robots could also be enhanced with interactive chatbot, similar to the WHO WhatsApp auto responder, which has the potential to reach 2 billion people.

## Blockchain technology in digital health systems

Blockchain technology could enhance the security and privacy of digital health systems while resolving specific problems in pandemic management (e.g. blockchain in virtual pharmacies for panic buying prevention). Dispensing non-prescription medicines (e.g. paracetamol, cough syrup), medical items (e.g. face masks) or household essentials (e.g. disinfectants) through low-tech dispensing machines is a quick and low-cost solution for 24-h available distribution with minimal human interaction. But as we have seen during the initial stages of the COVID-19 pandemic, people behave irrationally during crises and panic buying could lead to just a few people emptying the dispensing machines. In many countries, shops place restrictions on essential items, to preserve stocks from panic buying. Given that virtual pharmacies are a low cost, low human interaction solution, such restrictions need to be implemented through a technology that enables purchase control and monitoring. Panic during the COVID-19 pandemic was only a temporary issue. Therefore, for resolving short-term issues, purchase control mechanisms should either be low cost or enhance other functions of the systems (e.g. security and privacy).

Blockchain technology[33] has been in existence for some time. Open and public blockchain provides security without access control, serving as the transport layer for applications to be added to a network without the separate approval of trust. Design principles have already been proposed for 'tamper-resistant gathering, processing, and exchange of IoT sensor data' [67] privacy-preserving blockchain-based sensor data protection systems (SDPS). One of the use cases for evaluating the SDPS is IoT in a pharmaceutical supply chain.

We demonstrate in Fig. 8 how blockchain could operate in digital health systems.

We designed the diagram in Fig. 8 with data from this study, using the 'draw.io' open access web software.[34] There are permissionless (e.g. bitcoin) and permissioned blockchains (e.g. Facebook libra). The main difference for digital health systems is that permissioned blockchains are vetted by the network owner, whereas permissionless blockchains depend on anonymous nodes to validate transactions. In digital health systems, permissionless blockchains have advantages in preventing powerful actors in making decisions that favour individual groups at the expense of others.

Permissioned blockchains on the other hand are regulated by the authority of a specific ecosystem, usually a consortium, and present opportunities for creating interconnected systems. Such interconnected system could be designed by a consortium of healthcare providers, with patients' data belonging to the blockchain, and patients identified through unique ID providing both security and privacy of the patient. A patients' information sharing marketplace could prevent information blocking, and if cyber-attack results in information blocking, permissioned blockchains would present two unique advantages. Firstly, anyone in the medical consortium could check how and when information and transactions are happening. Secondly, blocking the information would change the hash on which the unique ID is based. Considering the above, the advantages for digital health systems are that patients can transmit personal records without the risk of tampering because blockchains are immutable and traceable. Apart from patients' security with blockchain permissions required for patient's medical data sharing, there are advantages in preventing fake or illegal sale of medication and with checking points of origin.

## Artificial intelligence for automatic diagnosis

Artificial intelligence systems are already in use for automatic diagnosis for conditions unrelated to COVID-19 (e.g. Zhongshan Ophthalmic Eye Center, China), and abnormal respiratory pattern classifiers are already in development for large-scale screening of people infected with COVID-19 [68]. As previously discussed, IoT connected thermal cameras could feed data into artificial intelligence systems for temperature screening in busy transport areas such as airports or train stations [45]. Such real-time data streams can support and enhance pandemic-risk management, but only if open protocols are set for safe and seamless communication between devices [44]. Unnecessary fragmentation of data restricts knowledge on trends and decisions on containment [44].

### Intelligent diagnosis, monitoring and supervision

Intelligent-assisted diagnosis, monitoring and supervision can be conducted relatively simply, and lessons can be learned from existing live tracking systems.[35] Online questionnaires can be uploaded into a real-time cloud database, and models can be updated based on the latest real case medical data to increase accuracy with results integrated in mobile applications automatically. Diagnoses can be grouped and automatically placed into 'confirmed, suspected or suspicious', while simultaneously, patients can be classified into 'mild, moderate, severe or critical pneumonia' categories [9]. The automatic patient classification may provide questionnaire on severity of infection, while the patient diagnosis grouping provides reassurances on the likelihood of infection. Such reassurances are aimed at reducing the pressure on medical professionals by limiting the need for in-person interaction.

### Tracking cases of prolonged isolation

Existing systems (e.g. Gov.uk/NHS vulnerability register[36]) can be used to create a mechanism for supermarkets to distinguish between customers in prolonged or age-related isolation and customers who are essential workers and those who are not a priority for home delivery services. Since this (online self-registration) mechanism already exists, preservation of privacy can be based on the existing system. This could be connected with a community-managed app linking volunteers[37] to those who are self-isolating and need help shopping, for example.

### Tracking high-risk cases

Detection and mapping of high-risk groups that are not currently being treated by the UK's NHS could help identify the most in need of immediate support due to isolation measures. In addition to tracking, digital health apps need to focus on predicting high-risk cases. Predicting vulnerability and criticality in patients can be achieved using ML models perhaps with input from activity sensors.

### Mental health impact mitigation

The COVID-19 pandemic could trigger new and sometimes unexpected mental health problems. Some mitigation activities could include free meals or relief on financial challenges or free showers to promote outdoor sports during work breaks. Information stations, like reception desks, could promote expression of concerns and keep workspaces safe while addressing staff concerns and anxieties. Social distancing will create a significant impact on mental health, and mental healthcare support needs to adapt to addressing not only the feelings of financial loss but also the feeling of emotional loss, the loss of self-worth, loss of motivation and a loss of meaning in daily life.[38]

## Discussion on gaps in scientific research

Our search of historic and recent scientific records on pandemics and new forms of data, on the Web of Science Core Collection, resulted in very limited records. In some cases, we found only a single record. It is unclear if this result is caused by lack of funding or lack of interest in the past on this research topic. Although we used all data records from the Web of Science Core Collection, there could be gaps in the data we are using. It could be too soon for publications with COVID-19 to have been published—and we should compare how different will it be in a year's time. We plan to rerun these queries again to see how the results have changes. In future studies, we should be focused on publication times, and if a vaccine is identified in the next few months or years, we should investigate how many papers have been published, comparable with the increase in COVID-19 activity. Future studies should also look at other factors of the virus in literature.

In any case, we can only discuss the gaps and results from the data records we currently have. Here is a brief outline of the gaps in scientific research.

Currently, there are only five scientific publications on the Web of Science Core Collection under the search on 'COVID-19 and big data', and only 59 on 'pandemics and big data', using historical records since 1900 (to 2020). Search results on 'big data' alone produced 88,410 records. There are 0 results on 'COVID-19 and time-stamped data' and on 'COVID-19 and genomics data'. The search of historic records, from 1900 to 2020, on 'pandemic and time-stamped data' resulted in only 3 records and on 'pandemic and genomics data' in 43 records. The search on 'COVID-19 and high-dimensional data' resulted in 0 records, and the search of historic records on 'pandemic and high-dimensional data' resulted in only 2 records.

Spatiotemporal data: currently, there is only one scientific publication on the Web of Science Core Collection under the search on 'COVID-19 and spatiotemporal data' and only 24 historic records since 1900 (to 2020) under the search on 'pandemic and spatiotemporal data'. Scientific research is needed on using spatiotemporal data in pandemic management.

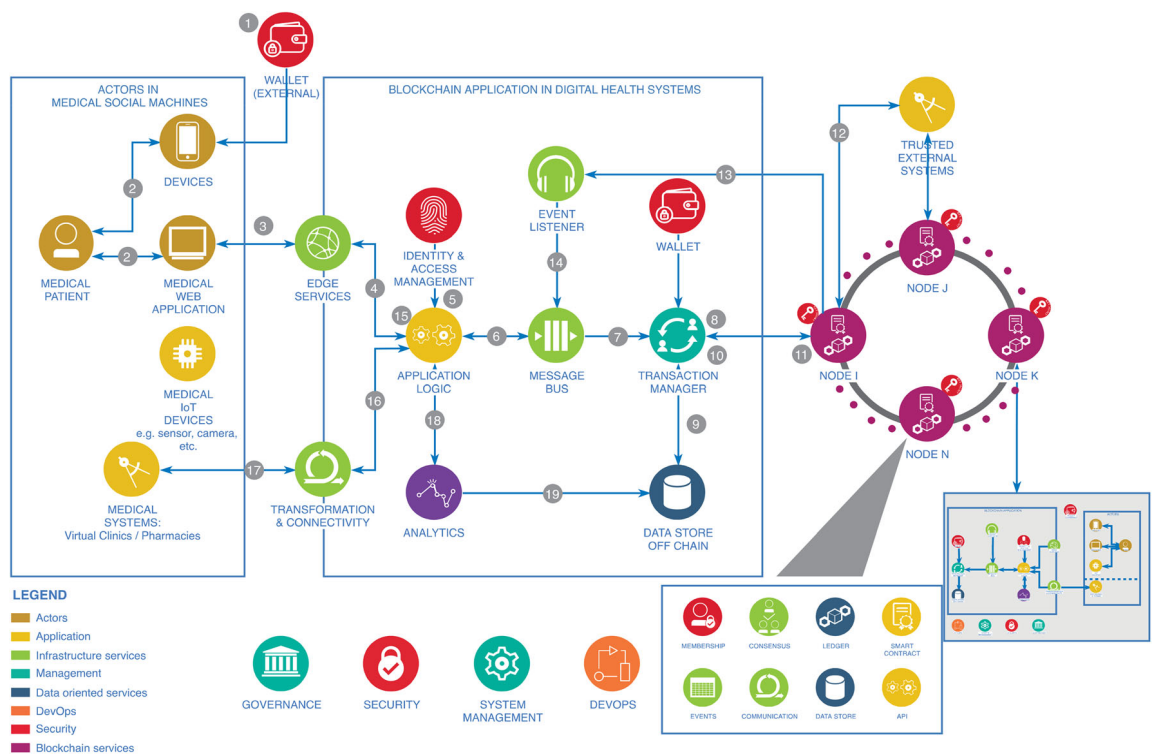

**Fig. 8** Blockchain in digital health systems integrated to COVID-19 virtual clinics

## Discussion on research objectives and research findings from the review

Following the example of previous studies [69, 70] related to predictive, preventive and personalized conclusions and expert recommendations, in this section, we review our research objectives and present our conclusions and expert recommendations.

The first objective of the paper was to review, compare and critically assess predictive, preventive and personalized digital technology responses to the COVID-19 pandemic around the world, with the aim of identifying a digital approach to COVID-19 and future global pandemics. Despite applying statistical software to analyse all of the existing literature on the Web of Science and other databases, the *conclusion* of the study is that we are currently not in a position to critically assess this technology and the best approach varies by country as it will be a function of wealth, smart phone usage and culture. Our *recommendation* on culture is that when choosing solutions, we must consider that in some societies like the USA, people tend to make decisions based on the individual ('shelter in place'), while others like China or Taiwan are more about the whole ('social isolation'), with Europe in the middle. Mirroring this, there are differing attitudes towards ceding privacy to the state or networks of corporations mentioned in the results.

The second objective and a point of interest in this paper was to review predictive, preventive and personalized interoperable digital healthcare solutions. The aim of this objective was to promote research that prevents failures from the past, where the separate design of digital health solutions has led to lack of interoperability. Hence, the review includes case studies that investigate different interoperable digital healthcare systems. A diverse set of non-contact digital healthcare technologies are investigated in case studies from different regions in the world. In our *conclusion*, we found no increase in personal data security risks from the integration of these digital technologies. What we did find was that some technologies have been used effectively in healthcare for decades (e.g. teleconferencing, mobile health, collaboration apps, virtual meetings, robotics, artificial intelligence), but we still failed in preventing the COVID-19 pandemic from spreading globally at a very fast rate.

Among many questions emerging from this review, in our *recommendation*, the following are considered to be important:

(1) How will we approach the problem of global healthcare strategies for predictive, preventive and personalized pandemic management?
(2) How will we deal with the lack of adoption of the reviewed non-contact technologies?
(3) How will we address the lack of digital investment and technological adaptation in predictive, preventive and personalized healthcare systems?
(4) How will we handle the resistance to change?

(5) How do we stop the problem of fake news about COVID-19 (and future pandemics) and cyber risks in personal data sharing with these established and regulated systems.

Apart from the multiple benefits identified for future global pandemic management, the review does not identify any new or unaccounted cyber risks from developing these interoperable digital healthcare solutions.

The third objective of the paper was to determine if the proposed new predictive, preventive and personalized technologies and data policies for mass surveillance linking personal data from health professionals to governments would actually address the pandemic crisis. The *finding* from the review paper is that public response is limited by a mistrust about data privacy and that this differs across countries, depending on the chosen public communication strategy.

Most countries' digital responses involve combinations of big data analytics, integration of national health insurance databases, tracing travel history from individual's location databases, code scanning and individual's online reporting. What is *missing* in the COVID-19 pandemic around the world is an integrated approach for digital healthcare management. In the review paper, we identified a plethora of tried and tested non-contact predictive, preventive and personalized digital healthcare concepts (e.g. mHealth, virtual clinics) that could help with *controlling* and *preventing* future pandemics. Mass surveillance linking personal data from health professionals should *not* be collected and used by governments but to the non-contact digital health technologies. The surveillance approach promoted by tech companies, where data is processed on individual devices and only anonymised data is shared with approval of the device owner, *seems safer* than some proposals for creating national databases. The *concerns* that big tech companies would abuse our personal data are in some cases *exaggerated*. Such personal data is already stored in our mobile devices and shared on social media sites, in mobile health apps, wearable devices or with our GP surgeries, in compliance with current data privacy regulations. Hence, *focus* should be placed on secure integration of this data, which is already shared, for the purposes of COVID-19 and future pandemics detection, monitoring and management. Such integration would create *numerous benefits* for the healthcare system, and we should anticipate an increased usage of these existing methods for sharing personal data.

Many new users would be *unfamiliar* with how mobile data is analysed on our secured individual devices and how anonymised data is shared. This would firstly trigger *fear* of personal data abuse; secondly it would *expose* users unfamiliar with data privacy to agreeing for their data being shared without understanding the data privacy policies. Hence, in times when mass personal data surveillance is *necessary*, a mass information campaign should also be focused on individual *understanding* of these new surveillance systems. This could be an *opportunity* for governments to enable tech companies to learn from the COVID-19 and develop predictive, preventive and personalized tools that could *prevent* or *minimize* the effect of future pandemics. We should *avoid mistakes from the past*, when we did not prepare and learn lessons from previous pandemics, that are reviewed in this paper.

## Conclusion

The most visible narrative from this review paper is that global pandemics require global response, and the world has failed to learn from previous pandemics. While big tech companies (Google and Apple) are promoting a unified global approach, some governments are promoting regional approaches. While regional approaches can work in localized geographical areas, the abilities of developing countries to adopt similar regional systems are questionable. Most of the required predictive, preventive and personalized technologies (e.g. mobile phones, mobile apps, low-cost connected devices (IoT) and Internet connection) are present in developing counties. But the abilities of some developing countries to develop regional predictive, preventive and personalized pandemic monitoring systems are questionable. Especially when considering the required speed for building such systems and the need to operate at low latency, this leads to the inevitable conclusion that such systems should be global and in place ready to be activated in cases of pandemics. Adding to this, the review study concludes that such global systems for pandemic management should also be designed as interoperable and coupled with existing digital non-contact healthcare systems.

To contribute to the design of such predictive, preventive and personalized interoperable and coupled digital non-contact healthcare systems, this review paper presents a survey of the Internet of Things and social machine literature as well as the overlap with the COVID-19 pandemic management. It continues with a survey of technology-driven mitigative measures and suggests some possible applications of technology in the COVID-19 pandemic management.

The review applied statistical software to review and analyse existing databases and concluded that we still lack the data to critically access what the best approach is to integrate digital technologies in healthcare and pandemic management. The best approach probably varies by country as it will be a function of wealth, smart phone usage and culture.

The review was enhanced with case studies on predictive, preventive and personalized interoperability of digital healthcare solutions and found that some technologies have been used effectively in healthcare for decades, and the risk of personal data exposure was not enhanced by these technologies. But these technologies failed in preventing the COVID-19 pandemic from spreading globally at a very fast rate. These technologies have the potential to prevent rapid spread of pandemics and even contain pandemics. Hence, future efforts should be placed on predictive, preventive and personalized pandemic management strategies and pandemic preparedness. The world has contained pandemics in the past. Many regions have failed to use the technologies described in this report, or in many cases even to learn from past successes, and have failed to manage this pandemic. This could change, and since the data is currently incomplete, we will run this analysis again in a year and see how it has changed and how it compares with previous pandemics. This study represents a snapshot in time and can be used in future studies for investigating different stages of the pandemic.

Finally, emerging mass population surveillance systems should be built in a predictive, preventive and personalized interoperable system, to promote advancement of non-contact digital healthcare concepts (e.g. mHealth), that could help controlling and preventing future pandemics. To promote such an interoperable approach, the new mass population surveillance system cannot rely solely on technology and need to operate as social machines. Social machines with large numbers of human-computer participants can provide predictive, preventive and personalized answers to different and complex pandemic monitoring problems.

**Acknowledgements** Eternal gratitude to the Fulbright Visiting Scholar Project.

**Availability of data and materials** all data file and materials are included in the article, or submitted as supplementing documents.

**Authors contributions** Dr. Petar Radanliev, main author; Prof. Dave De Roure, Prof. Max Van Kleek, supervision; Rob Walton, Dr. Rafael Mantilla Montalvo, Omar Santos, La'Treall Maddox, Stacy Cannady, review and corrections

**Funding information** This work was funded by the UK EPSRC [with the PETRAS projects: RETCON and CRatE, grant number: EP/S035362/1] and by the Cisco Research Centre [grant number DFR05640].

## Compliance with ethical standards

**Competing interests** The authors declared that they have no conflict of interest.

**Patient and public involvement** Patients and the public were not involved in this research.